\documentclass[aps,prb,twocolumn]{revtex4-1}
\usepackage{amsmath}
\usepackage{amssymb}
\usepackage{amsthm}
\usepackage{graphicx}
\usepackage{graphics}
\usepackage{bm}
\usepackage{hyperref}
\usepackage{rotating}
\usepackage{bbold}
\usepackage{color} 
 
\newcommand{\bra}[1]{\langle #1|}
\newcommand{\ket}[1]{|#1\rangle}

\def\beas{\begin{eqnarray*}}
\def\eeas{\end{eqnarray*}}
\def\bea{\begin{eqnarray}}
\def\eea{\end{eqnarray}}
\def\be{\begin{equation}}
\def\ee{\end{equation}}
\newcommand{\Ket}[1]{\vert  #1  \rangle}

\newcommand{\MatEl}[3]{\langle  #1 \vert #2 \vert #3 \rangle}
\newcommand{\Amp}[2]{\langle \, #1\, \vert\,  #2 \, \rangle}

\newcommand{\Avg}[1]{\langle  #1  \rangle}

\newcommand{\bpm}{\begin{pmatrix}}
\newcommand{\epm}{\end{pmatrix}}
\newcommand{\bmm}{\begin{matrix}}
\newcommand{\emm}{\end{matrix}}
\renewcommand{\epsilon}{\varepsilon}

\begin{document}

\title{Survival, decay, and topological protection in non-Hermitian quantum transport} 
\author{Mark S. Rudner$^1$, Michael Levin$^2$, and Leonid S. Levitov$^3$}
\affiliation{$^1$Niels Bohr International Academy and Center for Quantum Devices, Copenhagen University, Copenhagen, Denmark\\
$^2$ Department of Physics, James Frank Institute, University of Chicago, Chicago, IL, USA\\
$^3$ Physics Department, Massachusetts Institute of Technology, Cambridge, MA, USA
}
\date{\today}

\begin{abstract}
Non-Hermitian quantum systems can exhibit unique observables characterizing topologically protected transport in the presence of decay. 
The topological protection arises from winding numbers associated with non-decaying dark states, which are decoupled from the environment and thus immune to dissipation. 
Here we develop a classification of topological dynamical phases for one-dimensional quantum systems with periodically-arranged absorbing sites. 
This is done using the framework of Bloch theory to describe the dark states and associated topological invariants. 
The observables, such as the average particle displacement over its life span, feature quantized contributions that are governed by %are expressed through 
the winding numbers of cycles
around dark-state submanifolds in the Hamiltonian parameter space. Changes in the winding numbers at topological transitions are manifested in non-analytic behavior of %physical 
the observables. 
We discuss the conditions under which nontrivial topological phases may be found, and provide examples that demonstrate how additional constraints or symmetries can lead to rich topological phase diagrams.
\end{abstract}

%\pacs{}

\maketitle

% ----------------------------------------------------------------
%%%%%%%%%%%%%%%%% Introduction %%%%%%%%%%%%%%%%%%
 
\begin{figure}[t] 
\includegraphics[width=3in]{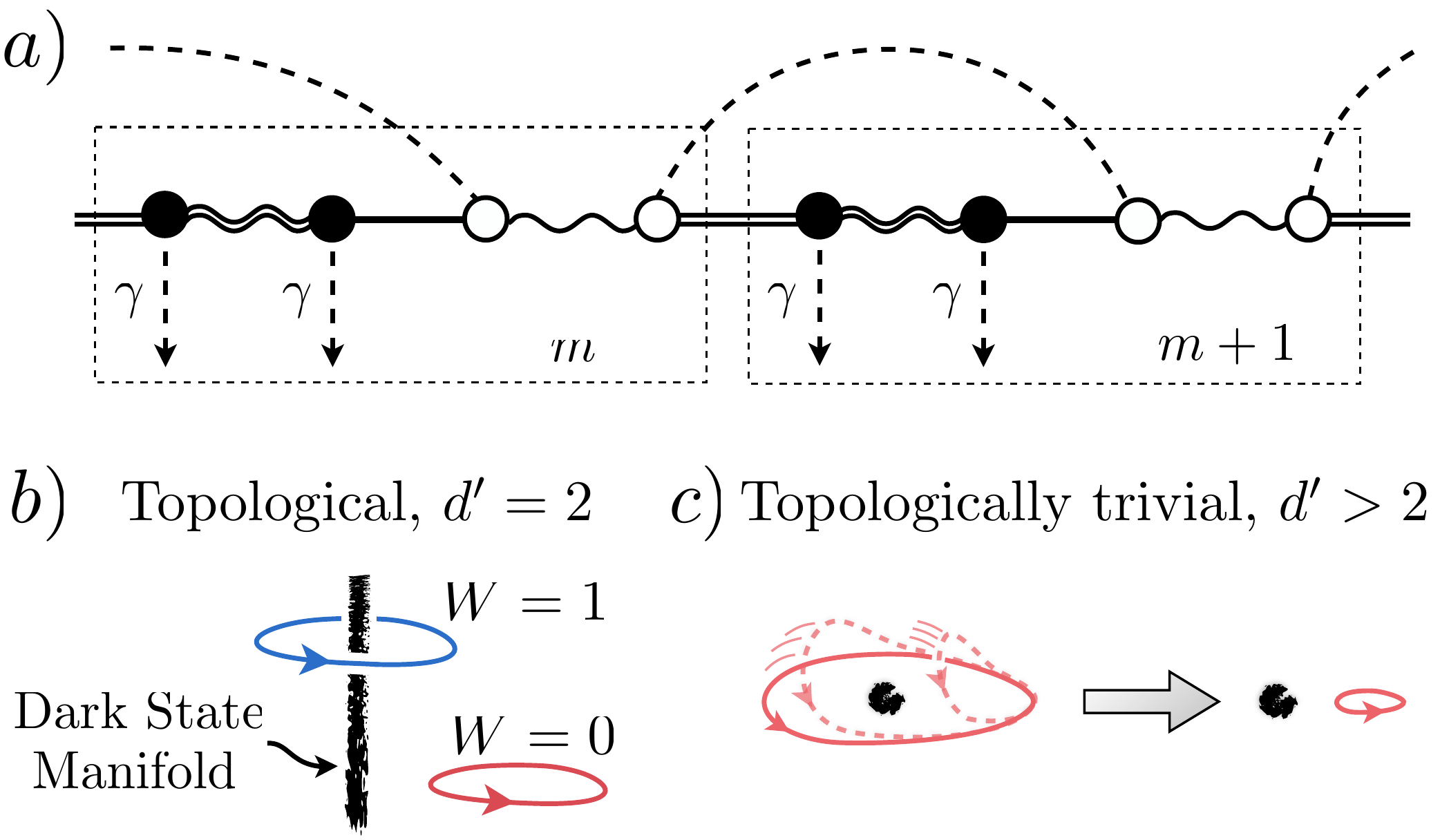} 
\caption[]{%Dimensional constraints on 
a) One-dimensional periodic non-Hermitian tight-binding system with $N$ sites per unit cell (dotted rectangles) with different hopping amplitudes shown by lines of a different kind. Within each unit cell, $M$ sites (filled circles) allow particle escape with the rate $\gamma$. 
Here the case $N = 4$, $M = 2$ is shown.
b), c) Topological classification of dissipative quantum dynamical phases. An $N$-band system's Bloch Hamiltonian [Eqs.\eqref{Hk_def},\eqref{HP}] defines a loop in the space of %can be represented by a closed loop in the space of 
$N \times N$ Hermitian matrices, parametrized by the crystal momentum $k$. 
%Quantization occurs
{Non-trivial topology arises} when there is exactly one decaying site per unit cell, $M = 1$. In this case the codimension of the dark-state %supporting 
manifold, $d' = 2M$,  is equal to two, see Eq.~(\ref{eq:codimension}). %has codimension $d'=2$.
Distinct phases are indexed by the winding number $W$, see Eq.\eqref{winding}.
%In the Fourier representation,
%Two Hamiltonians are in the same phase if their corresponding loops can be continuously deformed into one another without encountering a dark state for any $k$.
Quantization is absent for $M > 1$. % decaying sites per unit cell, 
%LL when the dark state manifold has codimension $d'>2$. In this case all loops can be contracted without obstruction.
{In this case,  since the dark state manifold has codimension $d'>2$, all loops can be contracted without obstruction.}
} 
\label{fig1}
\end{figure}

It was recognized recently that dissipative quantum systems
may exhibit unique transport phenomena of a {\it topological} character\cite{NH1D, NHDNP, Diehl2010, Kohmoto2011, Zeuner2015, Malzard2015, TonyLeeNonHerm2016}.
In particular, a new kind of quantized observable arises in a problem where particles with dynamics governed by a non-Hermitian Hamiltonian (or Lindblad master equation) can escape from the system whenever they visit a subset of sites on a periodic lattice\cite{NH1D} (see Fig.~\ref{fig1}). 
This observable is %, which characterizes particle transport in the presence of a decay, is 
given by the average displacement achieved by the particle over its life span in the system. %before escaping. % from the system.
%Although the quantization persists more generally\cite{MRNotes}, it is convenient to analyze the phenomenon in a Markovian setting in which particle escape is treated by including complex potentials on the decaying sites.
The displacement value was shown to have an interesting geometric meaning, namely that it is determined by a winding number defined in terms of the eigenstates of the non-Hermitian Hamiltonian that governs particle dynamics and decay\cite{NH1D}.
The topological transition associated {with 
%LL the changing of 
this winding number} was recently observed in an experiment using optical waveguide arrays\cite{Zeuner2015}.

Topologically distinct classes for this non-unitary evolution problem arise from the competition between survival and decay. 
% Particle loss (decay) plays a crucial role in generating {topologically distinct classes of evolution} for this non-unitary transport problem.
%LL , dissipation %coupling to the environment 
%LLplays a crucial role. 
Indeed, the quantity of interest -- the displacement achieved before escape -- can only be
%LL clearly 
unambiguously defined when particle dwell times in the system are finite. In this case, since each particle spends a finite dwell time inside the system, the displacement must vary continuously %under any
with system parameters. 
In contrast, discontinuous changes accompany transitions between different topological classes.  
Such transitions occur upon crossing boundaries in parameter space where one or more eigenstates of the system become completely decoupled from the environment and therefore can persist with infinite lifetimes.
The states with infinite lifetime, known as ``dark states,'' play a key role in a variety of phenomena in open quantum systems (see, e.g., Refs.~\onlinecite{EITReview, XuNature2009, LidarWhaleyDFS, Bardyn2013}). 
In our problem, these states capture the effects of long-time survival and, as we will see below, provide ``scaffolding'' for constructing topological classes. 
%%%%%%%%%%%%%%%%%%%%%%%%%%%%%%%%%%%%%%%%%%%%%%%%%%%%%%%%%%%%%%%%%%% 
%%%%%%%%%%%%%%%%%%%%%%%%%%%%%%%%%%%%%%%%%%%%%%%%%%%%%%%%%%%%%%%%%%% 

We note that our transport problem is distinct in several important ways from previously studied instances of geometric transport and Berry phases in dissipative systems, such as those discussed in Refs.~\onlinecite{Berry_phase_review,Avron,Sinitsyn}. One difference is that our problem, while dissipative, is quantum-mechanical, whereas the phenomena %dynamics 
analyzed in Refs.~\onlinecite{Berry_phase_review,Avron,Sinitsyn} are of a classical nature. Another difference is that Refs.~\onlinecite{Berry_phase_review,Avron,Sinitsyn} focus on geometric phase effects, in which the value of the Berry phase is in general non-quantized.
Here we are concerned with transport phenomena exhibiting quantization and topological protection.
In this regard one may seek a comparison with Thouless' quantized adiabatic transport\cite{ThoulessPump}.
However, while the Thouless pump is both quantum and topological, its dynamics are of a fundamentally {\it non-dissipative} nature.

Previously, non-Hermitian transport models of the type discussed in Ref.~\onlinecite{NH1D} have been employed to describe phenomena such as photon pumping in a cavity QED setting and nuclear spin pumping in open quantum dots\cite{NHDNP}, as well as to explore the classical-to-quantum transition in systems of interacting bosons\cite{Rapedius12}.
%\addMR{Other applications such as for describing coherent energy transport, e.g., in photosynthesis, can be envisaged as well.}\mpar{keep?}
Numerical investigations in Ref.~\onlinecite{NHDNP}, which studied nuclear spin pumping in spin-blockaded double quantum dots using a model based on that of Ref.~\onlinecite{NH1D}, revealed approximate quantization and the appearance of localized dark edge modes in a case where translational symmetry is broken.
These works along with the recent experiment in Ref.~\onlinecite{Zeuner2015} have raised interesting questions about the nature, robustness, and generality (e.g.,~applicability to other lattices) of this intriguing non-equilibrium topological phenomenon.

Motivated by these outstanding questions and with an eye on new experiments in a broad variety of systems -- atomic\cite{SorensenPrivateComm}, optical\cite{Zhao-PTSymm}, quantum optical\cite{NH1D}, and solid state\cite{NHDNP} -- here we generalize the model studied in Ref.~\onlinecite{NH1D} to arbitrary one-dimensional lattices.
We find that one or more winding numbers associated with the system's non-Hermitian Bloch Hamiltonian %non-dissipative part of the system's Hamiltonian 
index the distinct dynamical phases. 
Furthermore, we discuss how transitions between these phases generically give rise to non-analytic behavior of observables, giving clear signatures of the transitions.

Topology arises in our problem from a unique interplay between the Hermitian and anti-Hermitian parts of the dissipative system's Hamiltonian, with no analogue in conservative systems.
For a lattice with $N$ bands, the %coherent
{non-dissipative} part of the system's Bloch Hamiltonian defines a map from a circle, the one dimensional Brillouin zone, to the space of $N \times N$ Hermitian matrices.
The dissipative part of the Hamiltonian, in turn, defines a ``dark state manifold'' in this space, comprising the set of $N \times N$ Bloch Hamiltonians that support dark states.
As we will see, due to the necessity of avoiding dark states, %defines a manifold of obstructions
this manifold presents an obstruction in the $N \times N$ Bloch Hamiltonian space.
Generally, the dimension of the dark state manifold may not be high enough to break the topological equivalence between different Bloch Hamiltonian loops. %(see Fig.~\ref{fig:DimCounting}a). 
Topologically distinct loops may be found, however, if the codimension of the dark state manifold is equal to two. % (see Fig.~\ref{fig:DimCounting}b).
For the situation where no symmetries beyond the discrete lattice translation symmetry are present, we count the constraints associated with finding dark states and find that the codimension $d'$ of the dark state manifold is equal to $2M$, where $M$ is the number of decaying sites per unit cell (see Sec.~\ref{sec:DimCount}). %that this 
Thus, in the absence of symmetries, the codimension-two situation is achieved if and only if %occurs precisely when 
the system possesses exactly $M = 1$ decaying site per unit cell.
In this case, a single winding number characterizes the winding of the system's Bloch Hamiltonian around the 
dark-state 
%LL supporting 
manifold (Fig.~\ref{fig1}b). 

\begin{figure}[t] 
\includegraphics[width=3.2in]{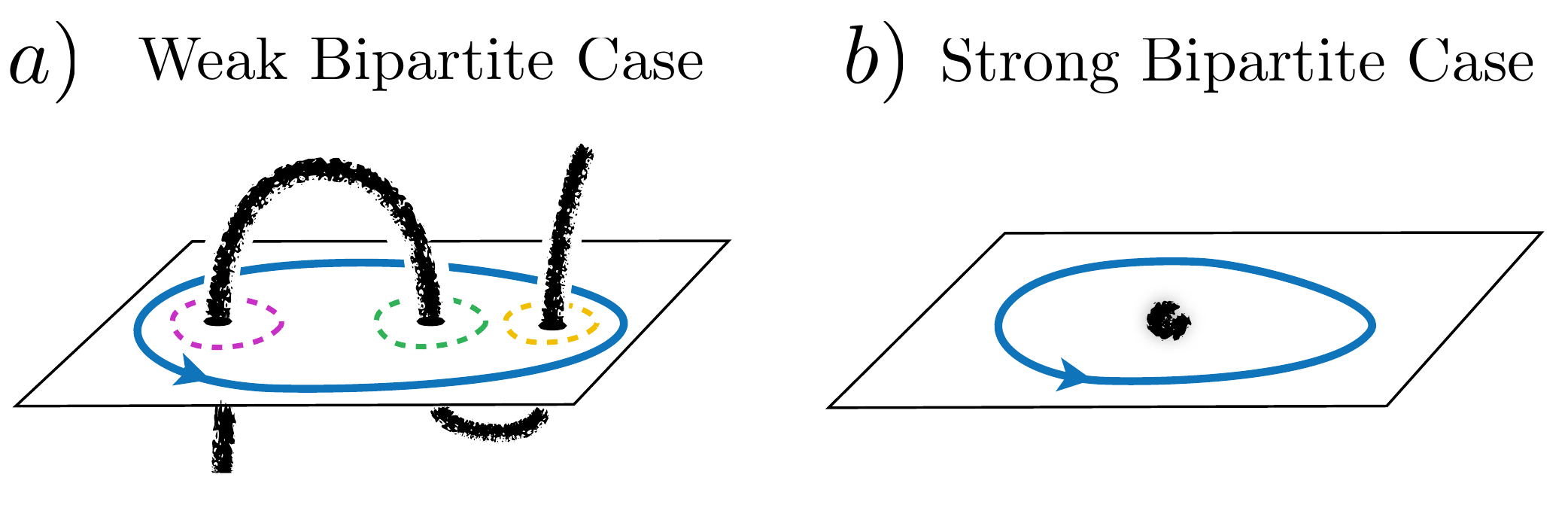} 
 \caption[]{The behavior of winding numbers is {schematically} illustrated for {Bloch Hamiltonians} with ``weak bipartite'' and  ``strong bipartite'' {constraints,} %Bloch Hamiltonians, 
see Eq.~\eqref{eq:weak_bipartite} and Eq.~\eqref{eq:strong_bipartite}.  
a) A ``weak bipartite'' constraint in the $M = 1$ case leads to additional windings (dashed circles).
%If the constraint is lifted, the single invariant which remains is equal to the sum of the individual winding numbers.
b) For $N = 2M$, with $M > 1$, a ``strong bipartite'' constraint restricts the space of allowed Bloch Hamiltonians, yielding a non-trivial classification.
{In this case topology is ensured by a chiral symmetry, analogous to the situation in 1D topological insulators.}} 
%% to generates a new topological classification where previously there was none.
%
%} 
\label{fig2} 
\end{figure}

The situation becomes more rich when additional symmetries or constraints are introduced.
In the presence of an additional ``weak bipartite constraint,'' defined below in Eq.~\eqref{eq:weak_bipartite}, the classification is broken down to a set of $N - 1$ independent winding numbers (see Fig.~\ref{fig2}a).
Were the symmetry to be removed, the system would be characterized by a single winding number as described above.
The value of the remaining invariant would then be given by the sum of the original $N-1$ winding numbers. %whose sum is preserved if the symmetry is lifted.
For a system with more than one decaying site per unit cell, $M > 1$, all Bloch Hamiltonian loops are equivalent if no symmetries are imposed.  However, in the presence of %where all loops are in principle equivalent, we show that  
a ``strong bipartite constraint,'' defined below, %provides sufficient restriction to 
the space of Bloch Hamiltonians may admit a non-trivial topological classification for $M > 1$ (Fig.~\ref{fig2}b).

It is interesting to compare and contrast the winding number that indexes the different dynamical phases in this problem with the invariants that appear in other familiar contexts {such as one-dimensional (1D) topological insulators\cite{SSH, Kitaev, Schnyder}. %, %electric polarization in crystals\cite{Resta}, 
There,} the presence of certain discrete symmetries is needed in order to obtain a nontrivial classification; in contrast, here no symmetries are needed.
Below we will show that, similar to electric polarization in crystals\cite{Resta}, the winding number is related to the Zak phases\cite{Zak} of our system's energy bands in the absence of decay.
Importantly, the winding number of the non-Hermitian problem is quantized and the physics of the problem gives distinct meaning to its different integer values.
In contrast, polarization is typically not quantized, and only its fractional part (i.e., its value modulo the lattice constant) is meaningful.

The remainder of the paper is organized as follows.
In Sec.~\ref{sec:Setup} we define the general class of non-Hermitian tight binding systems of interest. %whose dynamics we will consider.
In Sec.~\ref{sec:DimCount} we describe the geometry of the problem, and provide a dimension counting argument which reveals that the topological classification is generally trivial if the system possesses $M > 1$ %more than one 
decaying sites per unit cell.
In Sec.~\ref{sec:Class} we give the detailed topological classification for the nontrivial case $M = 1$, and define the winding number that indexes the distinct dynamical phases.
The role of additional symmetries is discussed in Sec.~\ref{sec:Symm}.
In Sec.~\ref{sec:Conseq} we discuss physical consequences of the topological classification. 
Our main results and conclusions are summarized in Sec.~\ref{sec:Summary}.

\section{Problem setup}
\label{sec:Setup}

Our system of interest is a general one dimensional translationally-invariant tight binding problem with an arbitrary number of sites per unit cell, $N$ (see illustration in Fig.~\ref{fig1}).
Single-particle hopping dynamics are described by the Hamiltonian
\bea
\label{H0}
H_0 = \sum_{m,m'}\sum_{\alpha,\beta} t_{m'-m}^{\beta\alpha}\ \ket{m'\, \beta}\bra{m\, \alpha},
\eea
where the (integer) indices $m$ and $m'$ label unit cells, and the Greek indices $\alpha, \beta =1, 2, \ldots , N$
%LL $\alpha, \beta \in \{1, 2, \ldots , N\}$ 
label the sites within each unit cell.
The hopping amplitudes $t_{m'-m}^{\beta\alpha}$ are translationally invariant, depending only on the displacement $m' - m$, and may cover an extended (but finite) range.
This tight binding problem is supplemented with a condition that $M$ out of the $N$ sites in the unit cell provide decay channels, allowing particles to escape from the system via coupling to an external continuum. %a particle may escape from the system 
%decay with a rate $\gamma$ occurs from a single site in each unit cell, $\alpha_0$. %a subset of sites containing $m$ sites from each unit cell, which are labeled $\alpha_1, \ldots \alpha_m$. 
As a matter of convention we choose a labeling such that the first $M$ sites, $\alpha = 1, \ldots, M$, correspond to the decaying sites.
Below we will show that, if no additional restrictions are imposed, the topological classification is nontrivial only for the case where there is exactly one decaying site per unit cell, $M = 1$. %in the situation with more than one decaying site per unit cell is trivial.

In this work we treat the decay within a Markovian framework, adding complex potentials on the decaying sites. %$\{\gamma_1, \ldots, \gamma_M\}$ 
%\addMR{Note that the assumption of Markovian decay is convenient, as it enables a simple formulation in terms of a non-Hermitian Hamiltonian; however, we expect the classification to hold for more general forms of decay.}
Note that, while the assumption of Markovian decay allows for a convenient formulation in terms of non-Hermitian Hamiltonians, we %do not believe that it is essential and 
speculate that the classification we develop is relevant in situations with more general forms of decay.
Extending the theory for non-Markovian decay presents an interesting challenge for future work. %\mpar{okay? need to say more?}
%%%%%%%%%%%%%%%%%%%%%%%%%%%%%%%%%%%%%%%%%%%%%%%%%%%%%%%%%%%%%%%%%%% 
%%%%%%%%%%%%%%%%%%%%%%%%%%%%%%%%%%%%%%%%%%%%%%%%%%%%%%%%%%%%%%%%%%% 

The state of the system $\ket{\psi}$ evolves according to the (non-Hermitian) Schr\"{o}dinger equation
\bea
\label{SE}i\frac{d}{dt}\ket{\psi} = H \Ket{\psi},\ H = H_0 - i\sum_{\alpha=1}^M \frac{\gamma_\alpha}{2}P_\alpha, %\Big[H_0 - i\frac{\gamma}{2}P_0\Big]\ket{\psi}, %\Big[H_0 - i\sum_{n=1}^{m}\frac{\gamma_n}{2}P_n\Big]\ket{\psi},
\eea
where %$P_0$ %
$P_\alpha = \sum_{m} \ket{m\, \alpha}\bra{m\,\alpha}$ 
is a projector onto the subspace of decaying sites of type $\alpha$, with $\gamma_\alpha$ the corresponding decay rate.  %decaying sites of type $\alpha_n$.
%with
%\bea
%\Gamma = \sum_x \gamma\, \ket{x\, \alpha_0}\bra{x\,\alpha_0}.
%\eea 
Due to the  translational invariance of the system, the problem (\ref{SE}) is most readily analyzed in the basis of Fourier modes, $\ket{k\, \alpha} = \sum_m e^{ikm}\ket{m\, \alpha}$.
%{\bf Introduce Fourier basis, ordered with $\alpha_0$ first.}
%Here, the problem can be analyzed in terms of a family of $N\times N$ Bloch Hamiltonians $\{H_k\}$, with $H_k = H_{0k} - i\frac{\gamma}{2}P_0$, parametrized by the $2\pi$-periodic crystal momentum $k$.
In the Fourier representation, the $2\pi$-periodic crystal momentum $k$ parametrizes a family of $N\times N$ Bloch Hamiltonians $\{H(k)\}$, defined through 
\be
\label{Hk_def} H = \oint \frac{dk}{2\pi}\, H_{\alpha\beta}(k)\,\ket{k\, \alpha}\bra{k,\beta}.
\ee
Decomposing $H(k)$ into its Hermitian and anti-Hermitian parts, we write $H(k) = H_{0}(k) - i\Gamma/2$, with
\begin{equation}
H_{0}(k) = \bpm %\epsilon_{0k} 
               \Delta(k) & v(k)^\dagger \\
		v(k)  &	h(k) \epm,\quad \Gamma = \bpm \bm{\gamma} & {0} \\ {0} & {0}\epm.
\label{HP}
\end{equation}
Here $\Delta(k)$ and $h(k)$ are $M \times M$ and $(N-M) \times (N-M)$ dimensional Hermitian matrices, respectively, $v(k)$ is an $(N-M) \times M$ dimensional rectangular matrix, and $\bm{\gamma}$ is a diagonal matrix with entries $\gamma_1, \ldots, \gamma_M$.
Physically, the matrices $\Delta(k)$ and $h(k)$ describe dynamics due to sublattice potentials and hopping within the decaying and non-decaying subspaces, respectively. Similarly, the matrix $v(k)$ describes the hopping between decaying and non-decaying sites.
In the topological classification scheme that follows, we will say that two systems %, with Hamiltonians $\{H_k\}$ and $\{H'_k\}$, 
are in the same phase if and only if their corresponding families of Bloch Hamiltonians $\{H(k)\}$ and $\{H'(k)\}$ %one family 
can be continuously deformed into one another without encountering a dark eigenstate for any $k$. 
%In this section, we find all the different phases for the $N \times N$ system.

\section{Constraints from dimension counting}
\label{sec:DimCount}
Before embarking on a detailed analysis, we first investigate the general constraints imposed by the geometry of the Bloch Hamiltonian space. 
From these considerations we deduce that the classification problem is trivial when there is more than one decaying site per unit cell ($M > 1$):
%That is, as long as 
if no additional constraints are imposed, all Hamiltonians with $M > 1$ belong to the same phase.

Consider an $N$-band system with $M \ge 1$ decaying sites per unit cell.
In this case, the $N \times N$ Bloch Hamiltonian $H(k)$ is specified by $N^2 + M$ real parameters, for each $k$: $N^2$ parameters define the $N \times N$ Hermitian part, $H_{0}(k)$, and $M$ parameters $\{\gamma_1,\ldots,\gamma_M\}$ specify the decaying part, $\Gamma$.
Because the decay rates $\gamma_\alpha$ must be strictly positive,  the set of Bloch Hamiltonians can be identified with $R^{N^2} \times (0,\infty)^M$. % (since $\gamma$ must be strictly positive).

The geometry of the problem becomes more interesting when we add the requirement that a particle initialized in {\it any state} %on any site 
must escape from the system with certainty in the infinite-time limit, i.e.,~that $H(k)$ must not support a dark state for any $k$.
A necessary and sufficient condition for a Hamiltonian $H(k)$ to have a dark state is that it has an eigenvector $y$ which 
vanishes on the entire decaying subspace: $y^T = (0, \ldots, 0, z_1, ... ,z_{N-M})$, in the representation introduced in Eq.\eqref{HP}. 
In terms of the parametrization in Eq.\eqref{HP}, such an eigenvector exists if and only if $z = (z_1, ... ,z_{N-M})^T$ is an eigenvector of $h_k$ and 
$v(k)^\dagger z = 0$. %where $\bf{0}$ is an $M\times 1$-dimensional column vector of zeros.
We conclude that $H(k)$ has a dark state if and only if $h(k)$ has an eigenvector $z$ satisfying 
$v(k)^\dagger z = 0$. 

While it is not easy to visualize the subset of Hamiltonians with dark states, here we will focus on its crudest
features -- in particular, its dimension. 
We note that the matrix relation $v(k)^\dagger z = 0$ stands for $M$ independent equations.
Satisfying all $M$ equations simultaneously imposes $2M$ real constraints on $H(k)$, since both the real and imaginary parts in each entry of $v(k)^\dagger z$ must vanish separately. 
It then follows that the subset of dark state Hamiltonians has codimension 
\be
\label{eq:codimension} d' = 2M.
\ee

This dimension counting is important because it gives strong constraints on when the Hamiltonians $\{H(k)\}$ can support a nontrivial classification. 
%The basic point is that 
Each family of Hamiltonians $\{H(k)\}$ corresponds to a closed loop 
in the Bloch Hamiltonian space. Two Hamiltonians $\{H(k)\}, \{H'(k)\}$ are equivalent if and only if the two corresponding loops can be deformed into one another without passing through the dark state subset. When the dark state subset has codimension greater than $2$, there is enough freedom that any loop can always be deformed around it, and contracted down to a point without obstruction. As an analogy, consider a loop in $R^3$ winding around a codimension $3$ subset, such as a single point (see Fig.~\ref{fig1}c). We conclude that the classification problem must be trivial when there is more than one decaying site per unit cell. 

In contrast, when $M = 1$, the subset of dark state Hamiltonians has codimension equal to $2$, and there can be topologically distinct ways for loops to ``wind'' around the dark state subset. Here, the analogy to think about is the winding of a loop in $R^3$ around a codimension $2$ subset, such as the $z$-axis (see Fig.~\ref{fig1}b). 
Loops with different windings cannot be continuously connected to one another without passing through the dark state subset and therefore correspond to distinct phases.

\section{Classification for $M=1$ case}
\label{sec:Class}
According to the preceding arguments, the only models which permit a nontrivial classification are those with 1 decaying site per unit cell ($M = 1$). We now explicitly identify all the possible phases for these systems, and construct a topological invariant which distinguishes them.

The first step is to find a way to parametrize the set $\mathcal{H}$ of $N \times N$ Bloch Hamiltonians %matrices
\begin{equation}
\label{HDecomp}
H(k) = \bpm \Delta(k) - \frac{i \gamma}{2} & v(k)^\dagger \\
		v(k)  &	h(k) \epm
\end{equation}
without dark states. % (here $\epsilon_0$ is a real parameter corresponding to the potential of the decaying sublattice).
Here, for the case $M = 1$, $\Delta(k)$ is a scalar and $v(k)$ is an $(N -1)\times 1$ dimensional column vector.
To this end, let $H$ be any such matrix. It follows from the lack of dark states that 
(a) each eigenvector $z_{n}$ of $h(k)$ (for $n = 1, \ldots, N-1$) satisfies $v(k)^\dagger z_{n} \neq 0$, and (b) the eigenvalues $\{\lambda_{n}\}$ of $h$ are non-degenerate.
The first implication is clear; as for the second, note that if any eigenvalue were degenerate, then 
by taking a suitable linear combination of eigenvectors one could construct a new eigenvector with $v(k)^\dagger z = 0$. 

Let $U$ be the $(N-1) \times (N-1)$ unitary matrix whose columns are the eigenvectors $z_{n}$, arranged in order of 
increasing eigenvalue, and with the phases of $z_{n}$ fixed by the condition that $v^\dagger z_{n}$ is real and 
positive. Then we have  (suppressing $k$ indices)
\begin{equation}
\label{UDecomp}
H = \bpm 1 & 0 \\ 0 & U \epm \cdot 
\bpm \Delta- \frac{i\gamma}{2} & \tilde{v}^\dagger \\
\tilde{v}   & \tilde{h} \epm \cdot 
\bpm 1 & 0 \\ 0 & U^{-1} \epm,
\end{equation}
where $\tilde{v}$ is real and positive, and $\tilde{h}$ is diagonal with entries 
$\lambda_{1} < \lambda_{2} < ... < \lambda_{N-1}$. 
Any $H$ without dark states can be written uniquely in this way; conversely, 
it is clear that any matrix of this form is a valid $H$ without dark states. 
Hence, the set $\mathcal{H}$ of $N \times N$ matrices without dark states can be parametrized in terms of (a) an $(N-1) \times (N-1)$ unitary matrix $U$, (b) a vector $\tilde{v}$ with real, positive entries, (c) a diagonal matrix $\tilde{h}$ with real entries $\lambda_{1} < ... < \lambda_{N-1}$, and (d) a complex scalar $\Delta-\frac{i\gamma}{2}$. 

It is now simple to find all the different phases for the $N \times N$ system: each Hamiltonian $H(k)$ corresponds to a closed 
loop in the space $\mathcal{H}$, so the problem of classifying different phases is equivalent to finding all topologically 
distinct closed loops in $\mathcal{H}$. Let us parametrize $H(k)$ by 
$(U(k), \tilde{v}(k), \tilde{h}(k), \Delta({k}) -\frac{i\gamma}{2})$. It is not hard to see that any closed loop $\tilde{v}(k)$ 
in the space $\mathbb{R}_+^{N-1}$ can be deformed into any other without obstruction, and similarly for $\tilde{h}(k)$ and 
$\Delta({k})-\frac{i\gamma}{2}$. Hence, the only component which could have any nontrivial topology is $U(k)$. As for this 
case, it is known that closed loops $U(k)$ in the space $U(N-1)$ are classified by their winding number $W$, defined by
\begin{equation}
W = \oint \frac{dk}{2\pi i} \partial_k \log \det U(k). %D_k), \quad D_k = \det(U_k).
\label{winding}
\end{equation}
Putting this all together, we conclude that the winding number $W$ gives a complete classification of the different possible 
phases of our system: two systems described by Bloch Hamiltonians $\{H(k)\}$ and $\{H'(k)\}$ are in the same phase if and only if they have the same winding number, 
as defined by Eq.\eqref{winding}.

%Note that translational symmetry played a key role in the description above.
Note that the existence of a Bloch Hamiltonian, parametrized by a well-defined crystal momentum $k$, depends crucially on translational symmetry.
Indeed, in the general case (see below for an exception where the topology is protected by an additional symmetry), the sharp distinction between phases %classification 
breaks down when translation symmetry is broken.
One way to see this is by imagining a periodic perturbation which doubles the unit cell.
Because the new unit cell contains more than one decaying site ($M > 1$), the classification becomes trivial: all Hamiltonians can be smoothly connected without encountering a dark state.
Alternatively, one may note that random on-site potentials %destroy the delicate phase relationships between wave function components on different sites which are responsible for the 
disrupt the destructive interference that is responsible for % maintains 
the dark state.
%Such sensitivity %should not be too concerning
%may not be too concerning, however, as many interesting 
Importantly, for many physical situations described by a model of the form above (see e.g., Refs.~\onlinecite{NH1D,NHDNP}),  %of this behavior can be found for 
the effective single particle hopping problem actually describes evolution between states in a many-body Hilbert space, rather than in real space.
In %some of these realizations, 
such contexts, disorder in the unit cell index may not be a natural perturbation. % to consider. %translation symmetry may be a much more robust feature.\mpar{Refs okay?}

Finally, to connect to familiar quantities, {we note that %it can be shown that 
the winding number in Eq.\eqref{winding} %can be shown to be equal to 
is proportional to} the sum of Zak phases over {\it all} $N$ bands of $H_{0}(k)$:
\be
W = -\frac{1}{2\pi}\sum_{n=1}^{N} \theta_n, \quad \theta_n = \oint {dk}\, \MatEl{k, n}{i\partial_k}{k, n},
\ee 
with
%$\partial_k \log \det(U_k) = \sum_{n=1}^{N} \MatEl{k, n}{\partial_k}{k, n}$, where 
$H_{0k}\Ket{k, n} = E_{kn} \Ket{k, n}$.
Here the phases of the eigenvectors $\ket{k, n}$ are chosen such that the component corresponding to the decaying sublattice is real and positive. %\mpar{appendix?}
While the Zak phase $\theta_n$ for any individual band $n$ is generally not quantized\cite{Zak}, the sum of Zak phases over {\it all bands} must be an integer multiple of $2\pi$.
The winding number, given by this integer, does not play a role in standard discussions of Zak phases because in the usual case the Zak phase is only meaningful mod $2\pi$, and different integer values of $W$ cannot be distinguished.
In contrast, in our problem the identification of the decaying subspace and the choice of unit cell uniquely fix a gauge, giving a physically distinct meaning to all integer values of $W$ that is reflected in observables such as the displacements investigated in Ref.~\onlinecite{NH1D}, {see also Sec.~\ref{sec:Conseq}.} % below.}

\section{Additional symmetries}
\label{sec:Symm}

As the dimension counting argument above demonstrates, imposing a ``no dark states'' requirement on non-Hermitian systems of the form described by Eqs. \eqref{H0} and \eqref{SE} allows for a classification into distinct topological phases. %occurs 
The inequivalence of different phases arises because of the restrictions imposed by the no dark states requirement on the space of allowed Bloch Hamiltonians. % imposed by .
If further restrictions or symmetries are introduced, additional topological phases, corresponding to new classes of non-contractible loops in the Bloch Hamiltonian space, may be generated. 
In principle, a wide variety of such additional symmetries could be introduced.
Here we give two examples for demonstration, and leave a further exploration of the role of various symmetries to future work.

In the first case we consider a ``weak bipartite'' constraint, which prohibits non-vanishing hopping matrix elements between
two decaying sites or two non-decaying sites in different unit cells. On the other hand, it allows for on-site energies and 
hopping between non-decaying sites {\it within the same unit cell} ~\cite{FootnoteBipartite}.
In this sense, this condition is weaker than a conventional bipartite constraint.
Note that any nearest-neighbor hopping model will satisfy this constraint. 
For the tight binding Hamiltonian given in Eq.\eqref{H0} the weak bipartite constraint reads: 
\be\label{eq:weak_bipartite} 
t^{\alpha\beta}_{m'-m} = 0
,\quad \alpha,\beta\le M\quad {\rm or}\quad \alpha,\beta> M
; \ \ \ m \neq m',
\ee
{where $m\neq m'$ applies in both cases.}

To obtain a classification in this case, we proceed similarly to above. First, we note that the weak bipartite constraint
implies that the $(N-1)\times(N-1)$ matrix $h\equiv h(k)$ in Eq.\eqref{HP} does not depend on $k$. As a result, the family of 
$(N-1)\times (N-1)$ unitary matrices $U(k)$ in the parametrization of Eq.\eqref{UDecomp} can be decomposed in terms of a constant ($k$-independent) unitary matrix $V$ and a family of diagonal unitary matrices $\{D(k)\}$: $U(k) = D(k)V$.
Here $V$ is used to diagonalize $h$, while $D(k)$ adjusts the phases of the eigenvectors to ensure that $\tilde{v}(k)$ in Eq.\eqref{UDecomp} is real and positive for each $k$. Given that $U(k)$ is of this special form, we can define $(N-1)$ different winding numbers -- one winding
number for each of the diagonal entries of $D(k)$. Each winding number describes how the corresponding element of $D(k)$ winds around the origin as a function of $k$~\cite{FootnoteWinding}.
If the bipartite constraint is lifted, the single invariant (\ref{winding}) that remains is the winding of the determinant of $D(k)$ --
i.e. the {\it sum} of these $(N-1)$ winding numbers. Thus we see that the additional restriction imposed by the weak bipartite constraint 
opens a richer topological structure in the system (see Fig.~\ref{fig2}a).

Next we consider systems satisfying a ``strong bipartite'' constraint. This constraint prohibits hopping matrix elements between two decaying or
two non-decaying sites, {\it whether or not they belong to the same unit cell}.  
On-site energy terms (any one of which can be regarded as a matrix element between a site and itself) are also prohibited. In addition, for a system with $N = 2M$ sites per unit cell, we require that exactly $M$ sites are decaying, and $M$ sites are non-decaying. 
This means that the systems we now consider will in general have more than one decaying site per unit cell. 
In terms of the tight binding model defined in Eq.~\eqref{H0}, the strong bipartite constraint reads superficially identically to Eq.~\eqref{eq:weak_bipartite},  
\be\label{eq:strong_bipartite} 
t^{\alpha\beta}_{m'-m}=0
,\quad \alpha,\beta\le M\quad {\rm or}\quad \alpha,\beta> M,
\ee
however with {the values of $m$ and $m'$ unconstrained, being allowed to be} %unconstrained values $m$ and $m'$ which can be 
{\it either equal or unequal.} 

According to the dimension counting argument, if no constraints were added, we would not expect to find more than one phase for $N > 2$ since then
there is more than one decaying site per unit cell.
However, the strong bipartite constraint severely restricts the set of allowed Hamiltonians, leading to a non-trivial topological classification for the case with equal numbers of decaying and non-decaying sites (see Fig.~\ref{fig2}d).
In fact, the strong bipartite condition implies that the Hermitian part of the Bloch Hamiltonian, $H_{0}(k)$, possesses a ``chiral'' symmetry defined by an operator $S$ that anticommutes with $H_{0}(k)$. 
In a representation where the amplitudes on decaying sites ($A$) are listed first, followed by the amplitudes on non-decaying sites ($B$), $H_{0}(k)$
and $S$ are described by 
\be
\label{Gamma}
H_{0}(k) = \left(\begin{array}{cc} 0 & H_{AB}(k)\\
H_{BA}(k) & 0\end{array}\right),\quad
S = \left(\begin{array}{cc} \mathbb{1} & 0\\
0 & -\mathbb{1}\end{array}\right),
\ee
Here $H_{AB}(k)$ describes the bipartite hopping and $\mathbb{1}$ stands for the $M \times M$ identity matrix.
%where $H_{0k}$ is the 
%where $L$ is the system size\mpar{Do this with $k$ instead?}.

From the theory of one dimensional topological insulators, it is known that a chiral symmetry of the form (\ref{Gamma}) leads to the existence of an integer-valued topological classification\cite{SSH, Kitaev, Schnyder}.
The value of the invariant is determined by the winding % (cite SSH, Shinsei, Schnyder, Kitaev).  
of $\det H_{AB}(k)$ as $k$ varies over the Brillouin zone. % in the Fourier-Bloch representation.
Importantly, in the case of the topological insulators, non-trivial topology arises from the requirement of maintaining a nonzero bandgap.
Here, a {\it different physical principle} (the requirement of complete decay) is responsible for the division into distinct topological classes.
Interestingly, in the special case of our non-Hermitian problem with the strong bipartite constraint, the nonzero bandgap and no-dark-state conditions coincide: both conditions are equivalent to the mathematical requirement that $\det H_{AB}(k) \neq 0$. 
Thus in this particular case the non-Hermitian problem follows the same classification as a one-dimensional topological insulator, albeit for a different physical reason.

Several important physical consequences can be inferred from this relationship between the topological insulators and the strong bipartite case.
First, given that the classification of insulators with chiral symmetry persists even without translational symmetry, the same robustness must hold for the non-Hermitian problem. Alternatively, one can establish this fact by directly generalizing the winding number (\ref{winding}) to disordered systems. Such a generalization can be obtained by considering a finite ring-like geometry and then analyzing the dependence of the Hamiltonian on the phase of a twisted-periodic boundary condition. % (see e.g.~Ref.~[{\bf Ref.}]).\mpar{what to cite?}

Another implication of this analogy is that phases with nonzero winding number must support ``dark edge modes'' localized near the ends of the system, when defined in a finite geometry with a boundary. 
To see this, recall that topological insulators with chiral symmetry feature robust localized edge modes with exactly zero energy.
% These 
Such states can be defined to live on only one sublattice or the other. 
Due to the boundary conditions at the edges, the zero-mode localized near one end will have its support on one of the two sublattices, while the zero-mode at the other end will have its support on the opposite sublattice.
For the edge which features a zero-mode with support entirely on the non-decaying ($B$) sublattice, we obtain dark edge states for the corresponding non-Hermitian problem.
The zero-mode localized at the other edge has its support on the decaying ($A$) sublattice and does not exhibit an enhanced lifetime.
%If we choose these eigenstates to be localized on the non-decaying ($B$) sites, we obtain dark edge states for the corresponding non-Hermitian problem.
We note, however, that these ``dark edge modes'' are only truly dark for an infinite system. 
For a system of finite length, hybridization between edge states at opposite ends of the chain, which live on opposite sublattices, endows the dark edge mode with a small decay rate that decreases exponentially in the system size.

Interestingly, such nearly dark states were observed in the experiment of Ref.~\onlinecite{Zeuner2015}, and in numerical investigations of a related in system in Ref.~\onlinecite{NHDNP}.
In the latter, a model of nuclear spin pumping in spin-blockaded quantum dots was mapped onto a non-Hermitian nearest-neighbor hopping problem of the form (\ref{SE}), restricted to a finite length chain.
The model satisfied the strong bipartite constraint but lacked translational symmetry: intracell hopping was considered uniform, occurring with an amplitude $u$, while intercell hopping amplitudes $v_m$ varied smoothly from nearly zero at the edges of the chain to a maximal value $v_{\rm max}$ near the center.  
For $v_{\rm max} < u$, the entire chain was in the trivial phase with winding $W = 0$.
Here nothing noteworthy was observed.
For $v_{\rm max} > u$, however, the central region of the chain entered the topological phase with $W = 1$, while the outer regions remained trivial with $W = 0$.
Thus two topological phase boundaries were formed.
As described above, in such a finite system we would expect to find an almost-dark state localized near one of the phase boundaries.
Indeed, the numerics showed that particles initialized near the rightmost phase boundary exhibited extremely long dwell times, characteristic of these nearly-dark states.
The dark states thus provide a potentially important experimental signature of the spin-pumping physics described by that model.

\section{Physical consequences}
\label{sec:Conseq}
\subsection{General considerations}
Given the existence of the topological classification outlined above, it is natural to expect that some observable properties of the system can be related to the associated winding number.
Such behavior was evident in Ref.~\onlinecite{NH1D}, where the average displacement of a particle initialized on one of the non-decaying sites in the $N=2$ case (two sites per unit cell) with nearest-neighbor hopping was found to be quantized, with its value equal to the value of the topological index.

More generally, consider any observable which can be written as an integral over the entire dwell time of a particle in the system,
\be
\label{Obs} X = \int_0^\infty\!\! dt\, \Avg{{\mathcal{O}}}.
%\Avg{\Delta \mathcal{O}} = \int_0^\infty dt\, \Avg{\dot{\mathcal{O}}},\ \dot{\mathcal{O}} = i[H,\mathcal{O}].
\ee
One may consider the observable for a particular initial state, or averaged over some distribution of initial states.
Such ``time-integrated observables'' have proven useful more generally in the study of dynamical phase transitions\cite{Hickey}.

When do we expect Eq.\eqref{Obs} to display non-analytic behavior?
For illustration, consider an initial state where the particle is localized on one of the non-decaying sites.  
At a transition point where $H_k$ supports a dark state for some $k$, the %or if all possible initial states are sampled uniformly, the 
average lifetime diverges due to the overlap of the initial state with the dark state: %the contribution of the dark state: %The lifetime of an initially localized particle diverges at a topological transition point due to its overlap with a dark state, 
$\bar{\tau} = \int_0^\infty dt\, \Amp{\psi}{\psi} = \infty$, see Ref.~\onlinecite{NH1D}.
Although for an infinite system the dark state itself has only infinitesimal occupation in the initial state, the contributions of nearby states with extremely long lifetimes give rise to a divergence via a van Hove-like singularity.
In such cases, the integral over time in Eq.\eqref{Obs} may not converge, giving rise to non-analytic behavior of $X$ %$\Avg{\Delta \mathcal{O}}$ 
at the transition.
Similar considerations apply for uniform sampling over all possible initial states.

In the classification arguments above we showed that, when there is exactly one decaying site per unit cell ($M = 1$), two Hamiltonians in different phases cannot be smoothly connected without crossing a dark state.
Therefore, we generically expect non-analytic behavior at the corresponding topological transitions.
%In analogy with familiar first order classical phase transitions, such as that at the liquid-vapor line,
Note that a path connecting two Hamiltonians in the {\it same phase} may also exhibit accidental crossings with the dark state manifold, thus giving rise to non-analytic behavior without a net topological phase change.
This behavior is analogous to that at a classical first order phase transition line which terminates at a critical point, as in the case of the liquid-vapor transition.
For cases with $M > 1$ and no additional symmetries, where the codimension of the set of dark-state supporting Hamiltonians is greater than 2, we generically do not expect non-analyticities to be encountered without fine tuning.
As a consequence, the breaking of translational symmetry, which can be viewed as an expansion of the unit cell, is expected to smooth out singularities in observables (see e.g.~Refs.~\onlinecite{NHDNP,Rapedius12}).

{
\subsection{Example: average particle displacements}
In the subsection above we argued that, quite generally, one may expect non-analytic behavior of time-integrated observables as parameters are tuned through topological transition points.
We now illustrate this behavior for the specific example of the average displacement achieved by a particle before decay, for $N \ge 2$ sites per unit cell. %, which can be written as:
In particular we will consider the situation where the displacement is averaged over a uniform distribution of all possible initial states.
Here we find that the averaged observable in fact contains a term which is directly given by the winding number itself, changing in quantized-height jumps across the transitions.}

{
As a first step, note that displacement is equal to the integral of velocity over time.
%Note that the net displacement can be represented in the form of Eq.\eqref{Obs} as
For a uniform distribution of initial states, the averaged displacement achieved over the particle's entire lifetime then takes the form of a time-integrated observable, Eq.\eqref{Obs}:
\be
\label{eq:displacement} \Avg{\Delta m} = \int_0^\infty\!\!\! dt \oint \frac{dk}{2\pi}\, \frac{1}{N}{\rm Tr}[ \partial_k H_{0}(k,t)],
\ee
where $\partial_k H_{0}(k, t) = e^{itH^\dagger(k)}\,\partial_kH_{0}(k)\,e^{-itH(k)}$ is the velocity operator in the (non-Hermitian) Heisenberg picture.
Here the choice of a fully random initial state is reflected in the uniform integral over $k$ and the trace over all $N$ bands.
}

{
Establishing a general relation between $\Avg{\Delta m}$ and the winding number $W$ in Eq.\eqref{winding} for $N > 2$ requires rather complicated additional technical machinery, and is beyond the scope of this work.
However, with the weak bipartite constraint discussed in Sec.~\ref{sec:Symm}, a direct, exact solution is possible (see Appendix).
Here we find that the expected displacement $\Avg{\Delta m}$, averaged with equal weights over all initial states, consists of a quantized piece given by the winding number $W$, on top of a non-universal background [see Eq.\eqref{eq:delta_m_N}]:
%For the case when all hopping amplitudes are real, the non-quantized background terms cancel out and we find
\be
\Avg{\Delta m} = \frac{W}{N} + (\textrm{analytic terms}). %,\quad \textrm{(weak bipartite case, $t^{\beta\alpha}_{m'-m}$ real)}.
\ee
Thus we see that topological transitions are reflected in the averaged displacement as quantized jumps of height $1/N$ each time the winding number changes by 1.
}

\section{Discussion}

In this work we have introduced a large class of one-dimensional 
%LL dissipative 
non-Hermitian systems which exhibit novel topologically protected transport phenomena. We developed a topological classification scheme for systems with periodically-placed absorbing sites, in which dark states whose wavefunctions vanish on the absorbing sites play a central role. Topology is captured by %and 
suitably defined winding numbers of the systems' non-Hermitian Bloch Hamiltonians around the dark state manifolds. 
Our results provide a framework for understanding the recent theoretical and experimental findings of topological phenomena in  dissipative systems of this type. They also provide guidance in the search for new topological phenomena in other dissipative systems.

The physical principle responsible for the classification is inherently linked to the presence of dissipation, through the requirement that any particle introduced to the system must escape in a finite time.
It therefore has no analogue in conservative systems.
We find that non-trivial topological classes are possible even when there no are symmetries apart from the translation symmetry of the lattice, provided that there is precisely one absorbing site per unit cell.
This stands in contrast to the situation for conventional topological bands in one-dimensional insulators\cite{Kitaev, Schnyder}, where particular discrete unitary and/or anti-unitary symmetries are needed to
obtain topologically distinct phases.

The models that we have analyzed are fundamentally defined on {\it discrete} lattices.
It is interesting to consider the consequences for {\it continuous} systems, such as the wave guide array used in the experiment of Ref.~\onlinecite{Zeuner2015}, where the description in terms of a tight binding lattice is only approximate.
Roughly speaking, a continuous system may be thought of in terms of the $N,M\rightarrow \infty$ limit for a family of discrete lattices with 
$M$ decaying sites per unit cell containing $N$ sites in total, with the ratio 
$N/M$ held fixed.
Based on the dimension counting argument developed in Sec.~\ref{sec:DimCount}, which indicates that non-trivial topology emerges only when $M = 1$, we anticipate that truly distinct topological phases may only emerge in a continuous system if decay occurs through an array of point-like sinks or ``single-channel'' absorbers. 

The behavior of such a continuum system, with the absorbers described by $M$ periodic arrays of delta functions in the system Hamiltonian,  will be quite different for $M=1$ and $M>1$. 
Classification for such a problem can be obtained by a direct extension of the above results, giving %a multitude of 
nontrivial classes for systems with $M=1$ and no nontrivial classes for generic systems with $M>1$ absorbers per lattice unit cell. Similar to the above discussion, systems with $M>1$ can exhibit nontrivial classes in the presence of additional symmetry constraints. However, a full understanding of the continuum non-Hermitian problem is currently lacking. The continuum regime thus remains an open and interesting topic for further study.

It is also natural to ask whether similar considerations can lead to a topological classification in higher dimensional systems. 
To date no such examples are known, but this is an interesting direction for future work.
In addition, it will be interesting to further explore the behavior of observables such as the average displacement and its higher moments in the general setting described above.
Such studies should go hand-in-hand with the search for new physical realizations of these interesting phenomena and their experimental implementations.

\label{sec:Summary}

We gratefully acknowledge interesting discussions with M. Rechtsman, M. Segev, and A. Sorensen. 
Support for this work was provided by the Villum Foundation and the People Programme (Marie Curie Actions) of the European Unions Seventh Framework Programme (FP7/2007-2013) under REA grant agreement PIIF-GA-2013-627838 [MR] and by the Center for Excitonics, an Energy Frontier Research Center funded by the US Department of Energy, Office of Science, Basic Energy Sciences under Award No. DE-SC0001088 [LL]. ML was supported in part by the NSF under grant No. DMR-1254741.

\appendix

\section{Expected displacement for systems with weak bipartite constraint}
In this appendix we derive a relation between the topological index $W$ in Eq.\eqref{winding} and the expected displacement achieved before decay, Eq.\eqref{eq:displacement}.
We focus on the case where the system possesses a weak bipartite constraint of the form in Eq.\eqref{eq:weak_bipartite}.
Specifically, we consider the case where the displacement is averaged over all possible initial states, where Eq.\eqref{eq:displacement} becomes
\be
\label{DeltamVel}\Avg{\Delta m} =\int_0^\infty\!\!\! dt \oint\!\frac{dk}{2\pi N}\, {\rm Tr}\left[ e^{itH(k)^\dagger}\!\left(\frac{dH(k)}{dk}\right) e^{-itH(k)}\right].
\ee
We will first demonstrate the approach (which differs from that used in Ref.~\onlinecite{NH1D}) for the case $N = 2$, and then show how it generalizes for arbitrary $N$.

To begin, suppose an operator $X(k)$ exists such that (with $k$-labels suppressed)
\be
\label{Xk}\frac{dH}{dk} = i\left[H^\dagger X - XH\right];\quad X \equiv \left(\begin{array}{cc} x_1& x_2\\x_3&x_4\end{array}\right).
\ee
Next we rewrite Eq.\eqref{DeltamVel} using Eq.\eqref{Xk} and the relation $\partial_t \left(e^{itH^\dagger }Xe^{-itH}\right)~=~e^{itH^\dagger}\left[iH^\dagger X - iXH\right]e^{-itH}$, which gives:
\bea
\!\!\!\!\!\!\!\!\!\!\!\Avg{\Delta m} = \frac{1}{N} \int_0^\infty\!\!\! dt \oint \!\frac{dk}{2\pi}\, \partial_t {\rm Tr}\left[e^{itH(k)^\dagger}\!X(k)e^{-itH(k)}\right]\!.
\eea
Because the integrand is a total derivative, the time-integral is simple and we are left with
\bea
\label{DeltaXk}\Avg{\Delta m} = -\frac{1}{N}\oint \frac{dk}{2\pi}\, {\rm Tr}\left[X(k)\right].
\eea
Here we used the fact that the upper limit of the time-integral gives zero, since $e^{-itH(k)} \rightarrow 0$ as $t\rightarrow\infty$.
All that remains is to show that such an $X(k)$ exists by finding its explicit form, and then to evaluate Eq.\eqref{DeltaXk}.

In the weak bipartite case that we consider, the Bloch Hamiltonian $H(k)$ takes the form %where $A$ sites only connect to $B$ sites: %, with long range hopping allowed:
\bea
\label{Hk}H(k) = \left(\begin{array}{cc} \epsilon & v^*(k)\\v(k)& \lambda\end{array}\right),\ v(k) = |v(k)|\,e^{i\phi(k)},
\eea
where $\epsilon = \epsilon_0 - i\gamma/2$ and $\lambda$ are {\it independent of $k$}.
All $k$-dependence appears in the off-diagonal matrix elements $v(k) = t_0 + t_1 e^{ik} + t_2 e^{2ik} + \cdots$.

Equations (\ref{Xk}) and (\ref{Hk}) define a set of 4 linear equations for the matrix elements $x_1, \ldots, x_4$:
\be\label{4x4}
\begin{array}{rcl}
\epsilon^* x_1 + v^*(k) x_3 - \epsilon x_1 - v(k)x_2 &=& 0\\
\epsilon^* x_2 + v^*(k) x_4 - v^*(k) x_1 - \lambda x_2 &=&-i\partial_k v^*(k)\\
v(k)x_1 + \lambda x_3 - \epsilon x_3 - v(k)x_4 &=&-i\partial_k v(k)\\
v(k)x_2 + \lambda x_4 - v^*(k) x_3 - \lambda x_4 &=& 0.
\end{array}
\ee
The linear system above is solved for
\bea
\label{2x2Soln}x_1 = 0,\quad && x_2 = -\frac{v^*(k)}{\gamma}\partial_k\ln |v(k)|^2,\\ x_3 = \frac{v(k)}{v^*(k)}x_2,\quad && x_4 = -\partial_k \phi + \frac{(\epsilon_0 - \lambda)}{\gamma}\partial_k \ln|v(k)|^2.\nonumber
\eea
Note that for any Hermitian $H(k)$ (i.e., for $\gamma = 0$), system (\ref{4x4}) has no solution.
Computing the trace, ${\rm Tr}[X(k)] = x_1 + x_4$, we find
\bea
\label{eq:2x2_soln}\Avg{\Delta m} %&=& -i \oint \frac{dk}{2\pi}\ x_1\\
&=& \frac12 \oint \frac{dk}{2\pi} \left[\partial_k \phi - \frac{(\epsilon_0 - \lambda)}{\gamma}\partial_k \ln|v(k)|^2 \right]\nonumber \\
&=& \frac12 \oint \frac{dk}{2\pi}\, \partial_k \phi.
\eea
The second term vanishes because $\ln |v(k)|^2$ is real, so $\oint dk\, \partial_k \ln|v(k)|^2 = 0$.
We recognize the integral in the second line of Eq.\eqref{eq:2x2_soln} as the winding number.
Note that, without the bipartite constraint, $k$-dependent terms could appear on the diagonal of $X(k)$, giving additional non-quantized contributions on top of the quantized piece found here.

How can this approach be generalized?
As discussed in Sec.~\ref{sec:Symm}, the weak bipartite constraint implies that the submatrix $h(k)$ of $H(k)$ in Eq.\eqref{HP} is {\it independent of $k$.}
Consequently, a single $k$-independent unitary transformation can be used to put $H(k)$ into the form
\be
H(k) = \left(\begin{array}{cccc} \epsilon & v^*_1(k) & v^*_2(k)  &\cdots\\ v_1(k) & \lambda_1 & 0 & 0\\v_2(k) & 0 & \lambda_2 &0\\ \vdots & 0 & 0 &\ddots\end{array}\right).
\ee
In this basis we seek a solution to Eq.\eqref{Xk}, where now $H(k)$ and $X(k)$ are $N \times N$ matrices.

For the specific case $N = 3$, Eq.\eqref{Xk} corresponds to
\begin{widetext}
 \be
 \left(\begin{array}{ccc}\epsilon^* & v^*_1 & v^*_2\\v_1 & \lambda_1 & 0\\v_2& 0 & \lambda_2\end{array}\right)
 \left(\begin{array}{ccc}x_1 & x_2& x_3\\x_4 & x_5 & x_6\\x_7 & x_8 & x_9\end{array}\right) -
 \left(\begin{array}{ccc}x_1 & x_2& x_3\\x_4 & x_5 & x_6\\x_7 & x_8 & x_9\end{array}\right)
 \left(\begin{array}{ccc} \epsilon & v^*_1 & v^*_2\\v_1 & \lambda_1 & 0\\v_2 & 0 & \lambda_2\end{array}\right) = 
 -i\left(\begin{array}{ccc}0 & \partial_k v^*_1 & \partial_k v^*_2\\ \partial_k v_1& 0 & 0\\ \partial_k v_2 & 0 & 0\end{array}\right).
 \ee
\vspace{0.1in}Through explicit calculation it is straightforward to find
 \be\label{sol3}
 X(k) = \left(\begin{array}{ccc} 0 & -\frac{v^*_1}{\gamma}\partial_k \ln |v_1|^2 & - \frac{v^*_2}{\gamma}\partial_k \ln |v_2|^2\\ 
        -\frac{v_1}{\gamma}\partial_k \ln |v_1|^2 & -\partial_k \phi_1 + \frac{A_1}{\gamma}& -\frac{v_1v^*_2}{\gamma(\lambda_2 - \lambda_1)}\partial_k\ln\frac{|v_2|^2}{|v_1|^2}\\ 
        -\frac{v_2}{\gamma}\partial_k \ln |v_2|^2 & -\frac{v^*_1v_2}{\gamma(\lambda_2 - \lambda_1)}\partial_k\ln\frac{|v_2|^2}{|v_1|^2}& -\partial_k \phi_2 + \frac{A_2}{\gamma}\end{array}\right),
 \ee
 with
 \bea
 A_1 &=& (\epsilon_0 - \lambda_1)\partial_k\ln|v_1|^2\ +\ \frac{|v_2|^2}{(\lambda_2 - \lambda_1)}\partial_k\ln\frac{|v_2|^2}{|v_1|^2}, \\
 A_2 &=& (\epsilon_0 - \lambda_2)\partial_k\ln|v_2|^2\ +\ \frac{|v_1|^2}{(\lambda_2 - \lambda_1)}\partial_k\ln\frac{|v_2|^2}{|v_1|^2}.
 \eea
%%\end{widetext}
Note the appearance of $\partial_k \phi_1$ and $\partial_k \phi_2$ on the diagonal of $X(k)$ in Eq.\eqref{sol3}.
Under the trace and integration over $k$, these contributions add up to give the winding number, $W = \sum_n \oint \frac{dk}{2\pi} \partial_k \phi_n$.
If specific initial states are targeted rather than averaging over all possible initial states, the individual winding number contributions corresponding to $\phi_1$ and $\phi_2$ may be revealed.

Examining Eq.\eqref{sol3}, we may hypothesize that $X(k)$ takes the following form for general values of $N$:
\be
X(k) = \left(\begin{array}{cccc}0 & a_{1} & a_{2} & \cdots\\ a^*_{1} & d_1 & b_{12} & b_{13}\\a^*_2& b_{21} & d_2 & b_{23}\\\vdots &b_{31}&b_{32}&\ddots\end{array}\right),
\ee
with
\bea
&a_n = -\frac{v^*_n}{\gamma}\partial_k\ln|v_n|^2,\quad b_{nm} = -\frac{v_n v^*_m}{\gamma(\lambda_m - \lambda_n)}\partial_k\ln\left\vert\frac{v_m}{v_n}\right\vert^2\nonumber\\
&\label{eq:d_n}d_n = -\partial_k\phi_n + \frac{(\epsilon_0 - \lambda_n)}{\gamma}\partial_k\ln|v_n|^2 + \sum_{m\neq n} \frac{|v_m|^2}{\gamma(\lambda_m - \lambda_n)}\partial_k\ln\left\vert\frac{v_m}{v_n}\right\vert^2.
\eea
By explicit calculation %, focusing separately on the first row, first column, diagonal, and off-diagonal elements, 
one can check that this solution indeed satisfies Eq.\eqref{Xk}. 
As above, we see that the appearance of $\partial_k \phi_n$ in the diagonal entries gives a contribution proportional to the winding number after integrating over $k$ and averaging over all initial states (i.e., taking the trace):
\bea
\label{eq:delta_m_N}\Avg{\Delta m} %&=& %\frac{1}{N}\oint \frac{dk}{2\pi}\,\sum_n d_k(n)\\
&=& \frac{W}{N} - \frac{1}{N}\oint \frac{dk}{2\pi}\, \sum_n \left[\frac{(\epsilon_0 - \lambda_n)}{\gamma}\partial_k\ln|v_n|^2 + \sum_{m\neq n} \frac{|v_m|^2}{\gamma(\lambda_m - \lambda_n)}\partial_k\ln\left\vert\frac{v_m}{v_n}\right\vert^2\right].
\eea
Note that when the hopping amplitudes are real, $|v_n|^2$ is even in $k$ and thus the second and third terms in the expression for $d_n$ vanish after integration over $k$, leaving {\it only} the quantized contribution from the winding number $W = \sum_n \oint\frac{dk}{2\pi}\partial_k \phi_n$. 
\end{widetext}

% \addMR{
% {\bf \noindent Menu}
% \begin{enumerate}
%   \item Read through paper
% %  \item Motivation: better understanding of phenomenon, explain $M > 2$
%   \item clarify connections with SSH?
%   \item importance of translation symmetry
% %  \item Relation to Kane isostatic lattice topological modes? Not at this time, connection not clear; no dissipation
% \end{enumerate}
% }

\begin{references}
\bibitem{FootnoteWinding}
Note that the definition $U(k) = D(k) V$ defines $V$ only up to an arbitrary phase convention.  
However, as long as the phase convention is chosen independent of $k$, the $(N-1)$ winding numbers associated with $\{D(k)\}$ are unique.
\end{references}
\end{document}